\documentclass[cits]{PoS}
\usepackage{graphicx}

\title{Epsilon regime calculations with reweighted clover fermions }

\ShortTitle{Epsilon regime calculations with reweighted clover fermions }

\author{\speaker{Anna Hasenfratz}\\
        
       Department of Physics, University of Colorado, Boulder, Colorado-80309-390,USA\\
       E-mail: \email{anna@eotvos.colorado.edu}}

\author{Roland Hoffmann\\
             Bergische Universit\"at Wuppertal, Gaussstra{\ss}e~20, 42219 Wuppertal,
Germany\\
\email{hoffmann@pizero.colorado.edu}}
\author{Stefan Schaefer\\
          Institut f\"ur Physik, Humboldt Universit\"at, Newtonstra{\ss}e~15, 12489
Berlin, Germany \\
\email{sschaef@physik.hu-berlin.de}}

\abstract{
We perform fully dynamical simulations at small quark masses by 
reweighting in the quark mass, 
calculating the weight factors stochastically. This approach avoids 
some of the technical difficulties 
associated with direct simulations. We find that the weight factors 
fluctuate only moderately on nHYP 
smeared dynamical Wilson-clover ensembles, 
and demonstrate that the overlap between the original and reweighted configurations is large 
both for short and long distance observables.
We  could successfully 
reweight $16^4$, (1.85fm)$^4$ volume 
configurations from $m_q\approx 20$ MeV to $m_q \approx 5$ MeV,
 and $24^4$, (2.77fm)$^4$ configurations from $m_q\approx 8$MeV 
to $m_q\approx 3$ MeV
quark masses, reaching the $\epsilon$-regime. 
Using the pseudoscalar and axialvector correlators we predict the low energy constants 
$\Sigma$ and $F$ and study their volume and mass 
dependence. 
}

\FullConference{The XXVI International Symposium on Lattice Field Theory \\
July 14 - 19, 2008\\
Williamsburg, Virginia, USA}

\begin{document}

\section{Introduction}

With efficient simulation techniques, 
 improved lattice actions and  increased computational
resources,
 essentially all  parameter values of lattice QCD, including
the point of physical quark masses, are accessible to direct simulation  today.
However, in the small quark mass regime the challenges
are still considerable:

\begin{itemize}
\item Large volumes are needed for the stability of the algorithms when
Wilson fermions are used 
\item Autocorrelation times increase dramatically towards the chiral limit 
\item Statistical fluctuations of fermionic correlators become difficult
to estimate since configurations with large contributions become rare
as small Dirac modes are more and more suppressed. 
\end{itemize}
A possible solution to counter these problems is to avoid generating
an ensemble with the fermionic weight of the light target quark mass
but instead simulate a heavier quark mass and reweight to the desired
ensemble. Simulational algorithms are more efficient at the larger
mass  and smaller volumes are sufficient from the algorithmic
point of view, while the autocorrelation time is controlled by the heavier
simulational quark mass. At a larger quark mass the region of small
Dirac eigenvalues is oversampled with respect to the target distribution,
 thus observables that receive large contributions there (e.g.
pseudoscalar correlators) will be better estimated. 

On the other hand reweighting will fail if the overlap between the generated
and target configuration ensembles is small, or if the weight factors
fluctuate strongly. One also has to overcome the computational problem
of  calculating the weight factors effectively. In two recent papers
\cite{Hasenfratz:2008fg,Hasenfratz:2008ce} we have developed a
 technique that solves the latter problem and demonstrated that
reweighting is possible and efficient on fairly large volumes between significantly
different quark masses. We have applied the technique to reach the
$\epsilon$-regime with Wilson fermions and predicted the values of
the low energy constants $\Sigma$ and $F$. In this paper we summarize the technique, 
giving further supportive evidence for the validity of reweighting.  
We also  briefly discuss the $\epsilon$-regime analysis,  
including the use of Wilson fermions in the $\epsilon$-regime and  the validity of the continuum chiral 
perturbation theory in our analysis.

An alternative approach to reweighting was presented in this conference \cite{Reweighting}.

\section{Reweighting in the quark mass\label{sec:lat}}

The numerical simulations for this project were done with 2 flavors
of nHYP smeared Wilson-clover fermions and one-loop Symanzik improved
gauge action\cite{Hasenfratz:2007rf}. We use the  tree-level $c_{SW}=1.0$ clover coefficient,
so our action is not fully $\mathcal{O}(a)$ improved, but we expect
that the $\mathcal{O}(a)$ corrections are small. We have generated
two sets of gauge ensembles, both at gauge coupling $\beta=7.2$.
The first set consists 180, $16^{4}$ configurations at $\kappa=0.1278$,
the second 154, $24^{4}$ configurations at $\kappa=0.12805$.

We set the lattice scale from the static quark potential, using $r_{0}=0.49$
fm for the Sommer parameter. On both configuration sets we found $r_{0}/a=4.25(2)$
, giving $a=0.1153(5)$ fm. With this value the physical volumes are
(1.85 fm)$^{4}$ and (2.77 fm)$^{4}.$ We connect the PCAC quark mass
to the renormalized one  using the pseudoscalar and axialvector
renormalization factors obtained with the RI-MOM method, 
$Z_{P}^{\overline{MS}}=0.90(2)$, $Z_{A}^{\overline{MS}}=0.99(2)$, 
 and estimate the renormalized quark mass in the $\overline{{\rm MS}}$
scheme at 2 GeV to be 22 and 8.5MeV on the 2 configuration sets, respectively. 

Starting from the original configurations one can explore a range
of quark masses in fully dynamical systems by reweighting the configurations.
The necessary weight factor is the ratio of the fermion determinants
that can be calculated stochastically, but one must take care not to introduce
significant statistical errors with the stochastic process. We apply
three methods, low mode separation, determinant breakup, and ultraviolet
(UV) noise reduction to control the statistical fluctuations. In both
the $16^{4}$ and $24^{4}$ ensembles we separate 6 low Hermitian
eigenmodes. In addition we break up the determinant to the product
of 33 and 60 terms for each $\Delta\kappa=0.0001$ shift in reweighting
on the $16^{4}$ and $24^{4}$ volumes, respectively.  

To control and
remove some of the UV noise we introduce an nHYP plaquette  pure gauge term in the
reweighted action with  coefficient  
 $\beta_{{\rm nHYP}}=6.0(\kappa-\kappa_{{\rm rew}})$.
This value
is so small that  within errors we do not observe its effect on the lattice
spacings or quark masses.
 It is worth emphasizing that including  the nHYP plaquette gives simply  an alternative gauge action 
and it does not introduce any systematical errors.

With the above outlined method we could reweight the $16^4$ configurations from $\kappa=0.1278$ to 
$\kappa=0.1281$, or approximately 5MeV quark mass, and the $24^4$ configurations from $\kappa=0.12805$ to 
$\kappa=0.12815$, or 3.8MeV quarks. The calculation automatically gives the weight factors at in-between 
$\kappa$ values and we analyzed both data sets at 4 different masses.

\begin{figure}
\begin{center}
\includegraphics[scale=0.6]{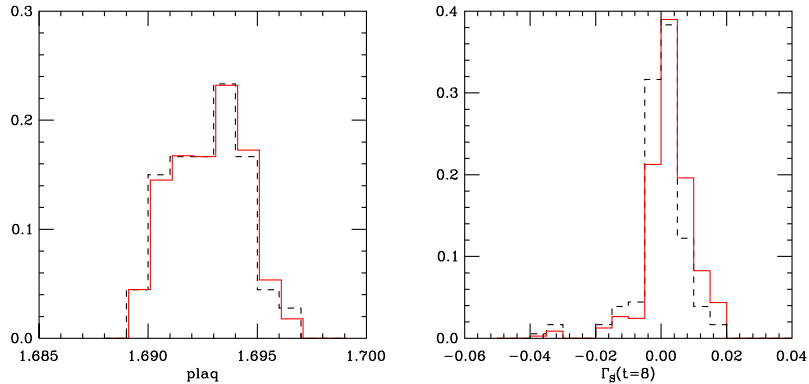}
\end{center}

\caption{The distribution of the plaquette (left panel) and scalar correlator
at $t=8$ (right panel) on the $16^{4}$ ensemble at $\kappa=0.1280$.
On both panels the dashed line is the original (partially quenched)
distribution and the solid lines correspond to the reweighted distributions.
The reweighting included the $\beta_{{\rm nHYP}}$ term as explained in the text.
\label{fig:overlap}}

\end{figure}

\begin{figure}

\begin{center}
\includegraphics[scale=0.45 ]{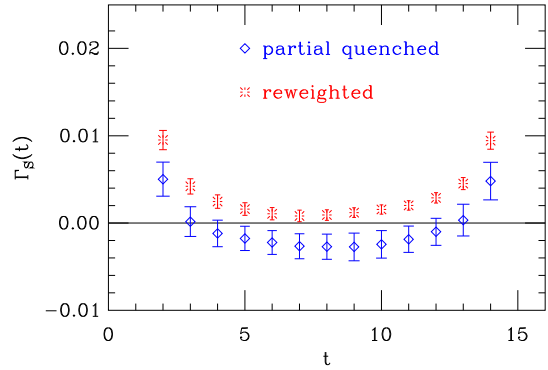}
\end{center}

\caption{The scalar correlator at $\kappa=0.1280$ for the partially quenched
and for the reweighted ensembles. \label{fig:The-scalar-correlator}}

\end{figure}

In addition to removing the UV fluctuations and thus reducing the fluctuations of the weight factors, 
the introduction of the
nHYP plaquette term also increases the overlap between the original
and target ensembles. The largest weight factors are pushed from the
edge of the plaquette distribution to the middle, where the statistical
sampling is better, and the effect is similar for other observables
as well. We illustrate this in Figure \ref{fig:overlap} with the
distributions of the plaquette and the scalar correlator at $t=8$.
All data correspond to the $16^{4}$ data set at $\kappa=0.1280$ ($\approx 9$ MeV quarks).
For both quantities the overlap between the reweighted and original
distributions is excellent, there is no sign that reweighting would
prefer region that is poorly sampled by the original ensemble. The
main difference between the reweighted and partially quenched scalar
data is the suppression of the negative contribution of the latter
one. This seemingly slight difference is nevertheless sufficient to
make the scalar correlator positive. In Figure \ref{fig:The-scalar-correlator}
we compare the partially quenched and reweighted, thus fully dynamical,
scalar correlators. Not only does the reweighted correlator stay positive,
its statistical errors are also significantly reduced compared to the
partially quenched one. The overlap between the reweighted and partially
quenched correlators for other mesons are similar, typically better,
than for the scalar one.

\section{$\epsilon$-regime analysis\label{sec:eps}}

In the $\epsilon$-regime the pion correlation length is large compared
to the linear size of the lattice, the light pseudoscalar mesons dominate
the dynamics. Nevertheless, in order to incorporate the massive modes
the volume has to be large compared to the QCD scale, or $L\gg F$. Chiral
perturbation theory predicts that at next-to-leading (NLO) order 
the meson correlators are quadratic  in  time and  depend  only on two low-energy constants, $\Sigma=\lim_{m\to0}\langle\bar{q}q\rangle$
and $F=\lim_{m\to0}F_{\pi}$. The explicit formulas have been calculated 
for the different meson correlators in continuum $\chi PT$, averaged over topological sectors, in 
\cite{Hasenfratz:1989pk,Hansen:1990yg}.
In the case of the Wilson lattice action 
one expects extra terms in the chiral expansion, but they show up only at 
next-to-next-to-leading order, so as long as the NLO formulas describe the data, 
the continuum expressions are sufficient. 

An important question to consider if Wilson fermions can be used at all in the $\epsilon$-regime. 
As the quark mass decreases in a large volume (p-regime) simulation,
the chiral symmetry breaking effects of Wilson fermions get large
compared to the mass, and that can create large lattice artifacts.
In practice the continuum limit has to be taken before the chiral
limit. The situation is different in the $\epsilon$-regime, where
the finite volume of the system creates an infrared cutoff even at
vanishing quark mass. This effect is well illustrated by the Hermitian
gap distribution. 
While in infinite volume one expects the
median of the gap to scale with the mass $\bar{\mu}=Z_{A}m_{{\rm PCAC}}$
\cite{DelDebbio:2005qa}, in the $\epsilon$-regime $\bar{\mu}$,
governed by the IR cutoff of the volume, remains finite while $m_{{\rm PCAC}}\to0$.
This is clearly the case in our simulations. While the quark mass changes by a  factor of 4 as we reweight from 
$\kappa=0.1278$ to $\kappa=0.1281$ on the $16^4$ ensemble, the median of the gap changes by about 40\% only
\cite{Hasenfratz:2008ce}.
One does not need a chiral
action to study the epsilon regime, though the explicit symmetry breaking
effects should be small compared to the inverse lattice size.

In this study we consider the pseudoscalar and the axialvector correlators. The former one 
is dominated by $\Sigma$, the latter by $F$, though both depend on the $\mathcal{O}(1)$
combination $m\Sigma V$ and  the $\mathcal{O}(\epsilon^{2})$ quantity  $1/(FL)^{2}$.  
The lattice correlators have to be multiplied by the renormalization
factors $Z_{P}^{2}$ and $Z_{A}^{2}$ to obtain the continuum ones, 
while in the product $m\Sigma V$ the quark mass can be expressed
in terms of the PCAC mass as $m=m_{{\rm PCAC}}Z_{A}/Z_{P}$. In our
analysis  we do a combined fit to the continuum pseudoscalar and axialvector correlators. 

The $\epsilon$-expansion formulas are systematic expansions in the
parameter $1/(FL)^{2}=\mathcal{O}(\epsilon^{2})$, but depend on the
$\mathcal{O}(1)$ quantity $m\Sigma V$. In our simulation we explore
the range $m\Sigma V\approx0.7\,-\,5.0$. Large values introduce large
NLO and NNLO corrections to the correlators, and at some point one
transitions into the large volume $p$-regime. Only by examining the
fit results can we decide what range of $m\Sigma V$ values
are acceptable in the $\epsilon$-regime.
The volume dependence is also an important issue to consider. On our smaller, (1.87fm)$^4$ volume, the
 $\epsilon$-regime expansion parameter is rather large, $1/(FL)^2\approx 1.45$, and
one expects large corrections to the NLO formulas. On our larger volume $1/(FL)^2\approx 0.46$, 
the NLO expansion is much more reliable.

Our philosophy is to fit both the small and large volume data with the NLO $\chi PT$ formulas
and study the quality of the fit and the change of the predicted low energy constants as
 the volume increases and quark mass decreases. 
\begin{figure}
\begin{center}
\includegraphics[scale=0.7]{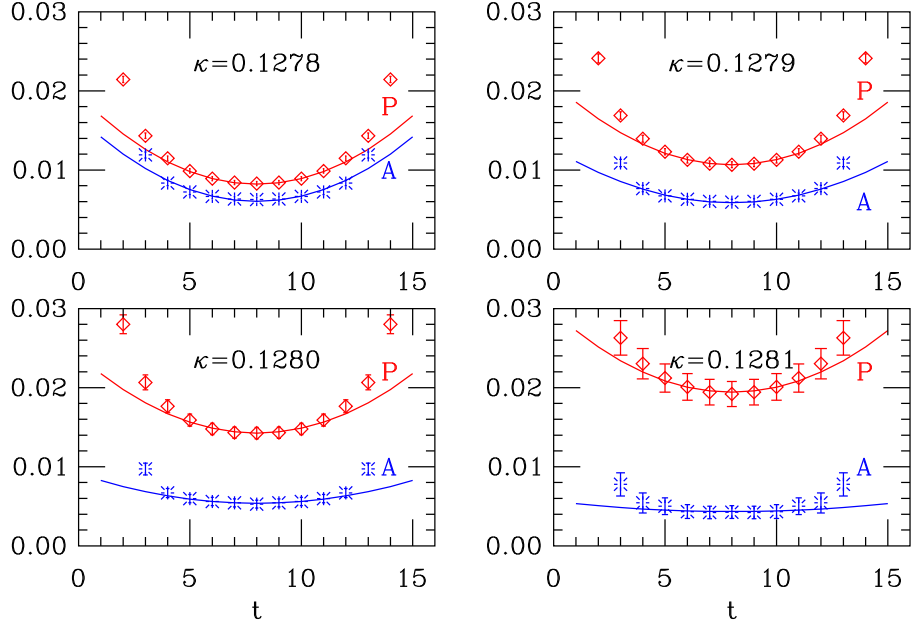}
\end{center}

\caption{The pseudoscalar ( red diamonds) and axialvector (blue bursts) lattice
correlators and the combined fit results on the $16^{4}$ data set.
The axialvector correlators are multiplied by the factor 50 to better
match the scale of the pseudoscalar.\label{cap:corr-16-1}}

\end{figure}

\begin{figure}
\begin{center}
\includegraphics[scale=0.7]{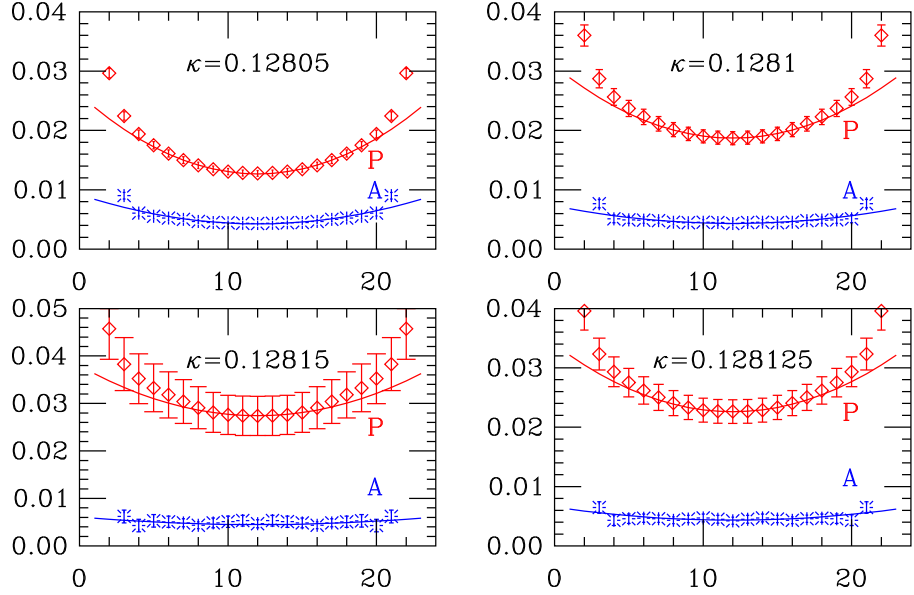}
\end{center}

\caption{Same as Figure 3 but on the $24^{4}$ data set
.\label{cap:corr-24-1}}

\end{figure}

\begin{figure}
\begin{center}
\includegraphics[scale=0.8]{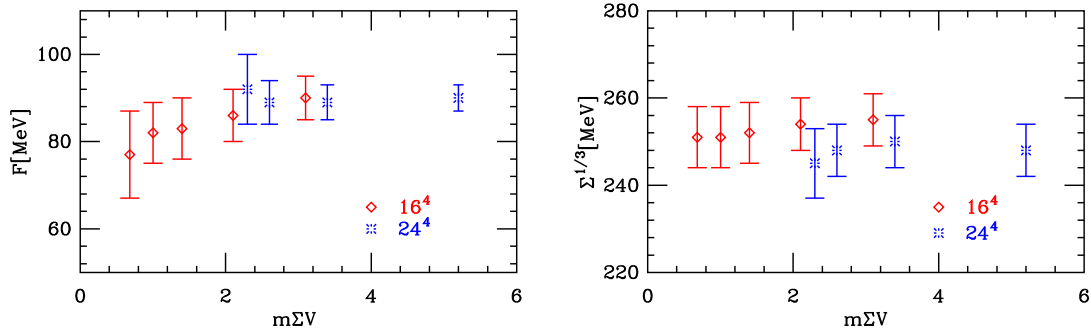}
\end{center}

\caption{The low-energy constant $F$ and $\Sigma^{1/3}$ as the function of
the parameter $m\Sigma V$, predicted by NLO $\chi$PT. \label{fig:Summary-plot}}

\end{figure}

The fits on the $16^{4}$ lattices are shown
in Figure \ref{cap:corr-16-1}, where we
plot both the pseudoscalar and axialvector correlators. 
The data are well described by the NLO formulas
at all four mass values. 
Figure \ref{fig:Summary-plot} shows the predictions for $F$ and $\Sigma$ as the function of 
$m\Sigma V$. While the condensate $\Sigma$ is basically constant, $F$ shows a slight drift as
the quark mass decreases. 

Our second data set is $24^{4},$ (2.77fm)$^{4}$, considerably larger.
Figure \ref{cap:corr-24-1}  shows
the combined fit and,  again, we find good agreement for all correlators.
The statistical errors are under control everywhere, though
they increase as the reweighting range increases .

The low energy constant $\Sigma$  on the larger volume is consistent with the
value obtained on the small volume (see Figure \ref{fig:Summary-plot}).
In case of $F$ the large volume data is independent of the quark mass and is
consistent with the small volume data where the two sets overlap. We cannot exclude the 
possibility that in the large volume data 
we see the breakdown of the $\epsilon$-expansion at large $m\Sigma V$, 
but we believe finite volume corrections are more likely
to explain the difference. Since the next-to-leading order corrections
to $F$ are over 10\% on the $16^{4}$ data set, we prefer using the
large volume data to arrive at our final prediction, \begin{equation}
F=90(4){\rm MeV,} \qquad \Sigma^{1/3}=248(6){\rm MeV}\,.\label{eq:results}\end{equation}
 The errors only include the statistical uncertainties. These values are consistent with 
other recent two-flavor computationsi\cite{Fukaya:2007pn,Jansen:2007rx}.

\section{Conclusion}

The data presented in this paper have been generated with moderate
computer resources. This was possible due to the good chiral properties
of the action which come at relatively low cost due to the simple
nHYP smearing procedure, and the effective reweighting that allowed
us to lower the quark mass even further. 
We have demonstrated good overlap for different observables between the original and reweighted data sets. 

With reweighting we were able to reach the $\epsilon$-regime with Wilson fermions
on 2 different volumes. Since the $\chi$PT expansion converges slowly in $1/(FL)^2$,
 our large
volume puts us into a good position and the comparison between the
$L/a=16$ and $L/a=24$ results shows that the finite volume effects
are under control. 

For the $\epsilon$ expansion to be valid, the parameter $m\Sigma V$
has to be $\mathcal{O}(1)$. Our data span the range 0.7 to 5.2 and
might go beyond the validity of the analytical expressions. An expansion
that connects the $\epsilon$ and $p$ regimes would be very useful
to control this aspect of the calculation.

Repeating the calculation at a smaller lattice
spacing would not be prohibitively expensive and could improve on
all of the above mentioned issues.

\section{Acknowledgement}

Most of the numerical work reported in this paper
was carried out at the kaon cluster on FNAL. We acknowledge the support
of the USQCD/SciDac.
 This research was partially supported by the US Department
of Energy and the Deutsche Forschungsgemeinschaft in the SFB/TR 09.

{\renewcommand{\baselinestretch}{0.86}
 \bibliography{lattice}
 \bibliographystyle{JHEP-2}}

\end{document}